\documentclass[%
 reprint,
showpacs,preprintnumbers,
 amsmath,amssymb,
 aps,
prd,
]{revtex4-1}

\usepackage{graphicx}% Include figure files
\usepackage{dcolumn}% Align table columns on decimal point
\usepackage{bm}% bold math
\begin{document}

\preprint{APS/PRD}

\title{Analytical Theory of Optical Black Hole Analogues}

\author{Suraj S Hegde}
 \email{shegde2@illinois.edu}
 \affiliation{ Research Education Advancement Program,Bangalore Assocication for Science Education,Bangalore,India\\
and\\ University of Illinois,Urbana-Champaign,Illinois,US}

\author{C V Vishveshwara}
\affiliation{ Former Senior Professor, Indian Institute of Astrophysics,Bangalore,India}

\date{\today}% It is always \today, today,

\begin{abstract}
We develop an analytical formalism for studying optical analogues of spherically symmetric black-hole spacetimes. We demonstrate the exact similarity between the electromagnetic wave equations in an inhomogeneous medium in flat spacetime and those in a general-relativistic curved spacetime respectively. The permittivity and the permeability of an inhomogeneous optical medium act as metric components of the analogue spacetime. Manifest properties of black holes like curved light trajectories follow directly from our formalism. Other black-hole phenomena like quasinormal modes can also be studied within this framework. We extend the above formalism and obtain an approximate analogue for the case of purely dielectric media. These optical black holes can be employed to investigate black-hole phenomena with table-top experiments.
\end{abstract}
\maketitle

%\tableofcontents

\section{\label{sec:level1}Introduction:}

Advances in Transformation Optics and Analog Gravity have led to systems that closely resemble those engendered by the general theory of relativity such as the black holes. Models that have come under the category of Analog Gravity \cite{visser} make use of sound waves in flowing media \cite{lorenci} like super fluids in Bose-Einstein Condensates \cite{hawking} \cite{volovik} to mimic certain aspects of a black hole . Such systems developed until now are not exact analogues of any real spacetimes with corresponding metric tensors. On the other hand, transformation optics has utilized the formalism of general relativity \cite{leon} to build gradient index static media in order to gain excellent control over the light paths in those media \cite{philbin}. Although much of these efforts in transformation optics have been directed towards developing technological applications, the results can be used effectively to investigate and construct models of optical analogues of gravity. In this context, we develop a general analytical formalism as the basis for optical black-hole analogues using static optical media. Our purpose is to generate in as much detail as possible, an optical analogue of the gravitational black hole and its manifestly characteristic phenomena, such as the existence of a one-way membrane, namely the event horizon, the bending of light rays, and the quasinormal modes. To achieve this end we must have: (1) A perfect light absorber which can act as the ‘event horizon’ and (2) A medium outside the ‘black hole’ that replicates the black-hole spacetime.
In the following sections we describe how a particular profile of permittivity and/or permeability of the optical analogue can lead to an ‘optical spacetime’ that has the black-hole metric and therefore reproduces the gravitational phenomena outside the ‘event horizon’.  Within the framework of our formalism, we further discuss the broadband omnidirectional light absorber developed by Narimanov and Khildishev \cite{narim} that serves exactly as ‘a perfect absorber’, or a black hole. We also show how one can, to a good approximation, simulate the black-hole spacetime with purely dielectric medium.

\section{\label{sec:level1}Electromagnetic Wave Equation in an Inhomogeneous Medium:}
The macroscopic electromagnetic field strength tensor in a continuous medium is defined as\cite{visser}
\begin{equation}
I^{\mu\nu}=Z^{\mu\nu\sigma\tau} F_{\sigma\tau}                                    
\end{equation}
                    
Where   $F_{0\mu}=-F_{\mu0}=-E_\mu$  ,$ F_{ij}=\epsilon_{ijk} B^k$,
      $ I^{0\nu}=I^{\nu0}=D^\nu $  , $ I^{ij}=\epsilon^{ijk} H_k$.                                                                                      
Here$ F_{\sigma\tau}$ is the field strength tensor in free space and $ Z^{\mu\nu\sigma\tau}$ is a fourth rank tensor whose components are the permittivity and permeability tensors. It is antisymmetric in the first two ($\mu,\nu$) and the last two ($\sigma,\tau$) indices.$ \epsilon^{ijk}$  is the Levi-Civita symbol. The roman indices span 1, 2, 3 while the greek indices range over 0,1,2,3.
The tensor Z is related to the permittivity and permeability tensors in the following way         
\begin{equation}
 Z^{0i0j}=-Z^{0ij0}=Z^{i0j0}=-Z^{i00j}=-\frac{1}{2} \varepsilon^{ij}    
\end{equation}
\begin{equation}   
 Z^{ijkl}=\frac{1}{2} \varepsilon^{ijm} \epsilon^{kln} {\mu_{mn}}^{-1}
\end{equation}
$\varepsilon^{ij}$  is the permittivity and $\mu_{mn}$ is the permeability.
Equation (1) is the succinct form of the two known equations:  $ D=\varepsilon E $  and  $ H=\mu^{-1} B $ .
For a source-free medium in flat space-time, the Maxwell’s equations involving sources are given by   
 \begin{equation}
(Z^{\mu\nu\sigma\tau} F_{\sigma\tau} )_{,\nu}=\mu_0 J^\mu  
 \end{equation}
In the absence of sources
\begin{equation}
 (Z^{\mu\nu\sigma\tau} F_{\sigma\tau})_{,\nu}=0        
\end{equation}                
For a curved space-time with metric $g^{\mu\nu}$, Maxwell’s equations are given by - 
\begin{equation}
 \frac{\partial}{\partial x^\nu }[\sqrt{-g}(g^{\mu\sigma} g^{\nu\tau}-g^{\mu\tau} g^{\nu\sigma})F_{\sigma\tau}]=0
\end{equation}
The similarity of eqs (5) and (6), suggest an analogy between an inhomogeneous medium and a curved spacetime, which is given by :
\begin{equation}
 Z^{\mu\nu\sigma\tau}=\frac{\sqrt{-\bar{g}}}{\sqrt\gamma}(\bar{g}^{\mu\sigma}\bar{g}^{\nu\tau}-\bar{g}^{\mu\tau}\bar{g}^{\nu\sigma})
\end{equation}
$\bar{g}^{\mu\sigma}$  is a function of the permittivity and permeability and is used with a bar on the top to distinguish it from the actual metric. Here $\gamma$ is the determinant of the spatial metric in the ambient curvilinear co-ordinate system(r, $\theta$,$\phi$) defined by $ \gamma_{\alpha\beta}= -g_{\alpha\beta}+ (g_{0\alpha} g_{0\beta})/g_{00} $ , where g is the metric of the ambient flat spacetime.  In spherical polar co-ordinate system on flat space-time,  $\sqrt\gamma=r^2\ sin{\theta}$ . 

One can directly formulate an optical spacetime at this stage itself, by defining an effective optical metric through the tensors $\varepsilon$ and $\mu$ ,which are the components of Z. However, we proceed to derive the wave equation for a spherically symmetric, magnetisable dielectric medium. With this equation as the basis one can easily get the Hamilton-Jacobi equation for the geometrical optics limit and also analyse the case of purely dielectric medium. The formulation of an optical spacetime will be achieved through the similarity between the wave equation in an optical medium and that in a curved spacetime respectively, which will formally bring out the analogy between the metric and the permittivity/permeability.

The field strength $F_{\sigma\tau}$ is written in terms of the four-potentials $A_\mu$, which are expanded in the vector spherical harmonics as follows\cite{cvv}
\begin{equation}
F_{\sigma\tau}=A_{\tau,\sigma}-A_{\sigma,\tau}
\end{equation}
\begin{equation*}
A_\mu=\sum_{lm}[(0,0,\frac{a^{lm}}{\sin\theta}\frac{\partial Y^{lm}}{\partial \phi},-a^{lm}\sin\theta\frac{\partial Y^{lm}}{\partial \theta}) 
\end{equation*}
Odd parity funtions
\begin{equation}
+(f^{lm}Y^{lm},h^{lm}Y^{lm},k^{lm}\frac{\partial Y^{lm}}{\partial \theta}, k^{lm} \frac{\partial Y^{lm}}{\partial \phi})] \end{equation}
Even parity functions.\\
$ a^{lm},f^{lm}, h^{lm}, k^{lm} $ are functions of  r and t.
$Y^{lm}(\theta,\phi)$ are scalar spherical harmonics.
Using the above in eq (5) gives, for odd parity functions the following equation
\begin{equation}
(\bar{\mu}_{\theta\theta}^{-1} a_{,r}^{lm})_{,r}-\bar{\varepsilon}^{\theta\theta}\frac{\partial^2 a^{lm}}{\partial t^2}-\frac{(l(l+1)}{r^2} a^{lm}=0
\end{equation}
And for even parity functions the following
\begin{equation}
(\bar{\mu}_{\theta\theta}^{-1} b_{,r}^{lm})_{,r}-\bar{\varepsilon}^{\theta\theta}\frac{\partial^2 b^{lm}}{\partial t^2}-\frac{l(l+1)}{r^2} b^{lm}=0
\end{equation}

Where $ b^{lm}$ is given by $ \bar{\varepsilon}^{\theta\theta}\bar{\mu}_{\theta\theta}^{-1} [(h^{lm})_{,0}-(f^{lm})_{,r}]=\frac{l⁡(l+1)}{r^2} b^{lm} $

These equations are similar to those developed for the gravitational case, by J.A.Wheeler and D.Brill\cite{Wheeler} and also by R.Ruffini, J.Tiomno, C.V.Vishveshwara\cite{cvv} (who include the source term) which are shown below 

\begin{equation}
(g^{rr}a_{,r}^{lm})_{,r}-g^{00}\frac{\partial^2 a^{lm}}{\partial t^2 }-\frac{(l(l+1))}{r^2} a^{lm}=0
\end{equation}
And for even parity functions it is
\begin{equation}
(g^{rr} b_{,r}^{lm})_{,r}-g^{00}\frac{\partial^2 b^{lm}}{\partial t^2 }-\frac{l(l+1)}{r^2}  b^{lm}=0
\end{equation}

These equations resemble the Regge-Wheeler equations for gravitational perturbations.
The similarity in equations (10), (11) and (12),(13) gives rise to similarity in a number of phenomena which depend on the form of the wave equation. One can formulate an ‘optical spacetime’ with an effective optical metric whose components are the permittivity and the permeability tensors.

$\bar{g}^{rr}=\bar{\mu}_{\theta\theta}^{-1}=\bar{\mu}_{\phi\phi}^{-1}$,
$\bar{g}^{00}=-\bar{\varepsilon}^{\theta\theta}=-\bar{\varepsilon}^{\phi\phi}$,
$\bar{\varepsilon}^{\theta\theta}\bar{\mu}_{\theta\theta}^{-1}=1$

Metamaterials can be used to realize such an optical spacetime. Such a system has been studied particularly for the Schwarzschild spacetime in \cite{chen} .

Choosing $\bar{\varepsilon}^{\theta\theta}=(1-\frac{L}{r})^{-1}$ and $\bar{\mu}_{\theta\theta}^{-1}=(1-\frac{L}{r})$,one gets an exact analogue of Schwarzschild black-hole spacetime, where r=L can be considered to be the location of the event horizon co-ordinate. The analysis of the gravitational bending of light and quasinormal modes will follow exactly as in the case of the gravitational case.

\section{\label{sec:level1}Wave equation in a non-magnetisable dielectric medium:}
Now consider the case of a non-magnetisable dielectric medium (with $\bar{\mu} =1$).This is of interest because of the simplicity of using only a dielectric medium for experiments.
  
With $\bar{\mu} =1$ equation (10) becomes
\begin{equation}
\frac{1}{\bar{\varepsilon}^{\theta\theta}}(a_{,r}^{lm})_{,r}-\frac{\partial^2 a^{lm}}{\partial t^2 }-\frac{l(l+1)}{\bar{\varepsilon}^{\theta\theta} r^2}a^{lm}=0
\end{equation}
To transform the equation to Regge-Wheeler form, substitute $a^{lm}=\frac{\psi(r,t)}{(\bar{\varepsilon}^{\theta\theta})^{1/4}}$ into the above equations to get
\begin{equation}
[\frac{\partial}{\partial r}(\frac{1}{\sqrt{\bar{\varepsilon}^{\theta\theta}}}\frac{\partial}{\partial r})-\sqrt{\bar{\varepsilon}^{\theta\theta}}\frac{\partial^2}{\partial t^2}-P(r)]\psi=0
\end{equation}

Here\\$ P(r)=\frac{l(l+1)}{\sqrt{\bar{\varepsilon}^{\theta\theta}} r^2}-\frac{5}{16}\frac{1}{(\bar{\varepsilon}^{\theta\theta})^{\frac{7}{2}}}(\frac{\partial{\bar{\varepsilon}^{\theta\theta}}}{\partial r})^2+\frac{1}{4}\frac{1}{(\bar{\varepsilon}^{\theta\theta})^{\frac{5}{2}}} \frac{\partial^2\bar{\varepsilon}^{\theta\theta}}{\partial r^2}$.\\
So,the effective spacetime metric can be written as follows $\bar{g}^{00}=-\sqrt{\bar{\varepsilon}^{\theta\theta}}$ ,
$\bar{g}^{rr}=\frac{1}{\sqrt{\bar{\varepsilon}^{\theta\theta}}}$

The function P(r) represents the effective potential, which in this case is not similar to that in (10). Therefore the behaviour of EM waves in the medium is slightly different but in certain approximation it will be similar to the gravitational case.
The derivatives of  $\bar{\varepsilon}^{\theta\theta}$ can be very small and hence we can neglect them in P(r).Further approximations depend on the specific form of $\bar{\varepsilon}^{\theta\theta}$  considered and also on the features to be studied using the analogue spacetime.

\section{\label{sec:level1}Discussions:}
We discuss the application and extension of the formalism in the case of purely dielectric medium which will involve some approximations. For magnetisable dielectric medium the procedure is exactly the same.
\subsection{\label{sec:level2} Schwarzschild Black-Hole analogue in dielectric medium}
Choice of $\bar{\varepsilon}^{\theta\theta}=\frac{1}{(1-\frac{L}{r})^2}$ gives us the analogue of the Schwarzschild spacetime. Substituting it into the wave equation gives
\begin{equation*}
\frac{\partial}{\partial r}[(1-\frac{L}{r})\frac{\partial}{\partial r}]\psi - (1-\frac{L}{r})^{-1} \frac{\partial^2 \psi}{\partial t^2} - \frac{(l(l+1))}{r^2} \psi+[\frac{l(l+1)}{r^3}L
\end{equation*}
\begin{equation}
+ derivatives  of   \bar{\varepsilon}^{\theta\theta}] = 0
\end{equation}

Here L represents the ‘Event Horizon’ of the analogue black hole. We are only interested in the physics of the phenomena happening outside the event horizon i.e. at $ r > L$. If L is sufficiently small, then outside the black hole we can ignore terms like $ L/ r^3 $ in the effective potential P(r).Also, if we consider a medium with its permittivity very slowly varying along r, we can ignore the terms with the derivatives of $\bar{\varepsilon}^{\theta\theta}$ .Therefore we have a simplified equation
 \begin{equation}
\frac{\partial^2 \psi}{\partial r_*^2} -  \frac{\partial^2  \psi}{\partial t^2} -(1-\frac{L}{r}) \frac{l(l+1)}{r^2} \psi
\end{equation} 
This is exactly similar to the gravitational case.Here we have used the Regge-Wheeler tortoise co-ordinate $ r_*=r+L\ln⁡(\frac{r}{L}-1)$.
The following are the two metric components of the effective optical spacetime
$ \bar{g}^{00}=-(1-\frac{L}{r})^{-1}$  , $\bar{g}^{rr}=(1-\frac{L}{r})$.
The above identification ensures that the optical spacetime in fact corresponds to the Schwarzschild with the event horizon located at r = L. Thus the optical system devoid of any magnetisable media can be used to simulate the black-hole spacetime, without the use of metamaterials.

\subsection{\label{sec:level2}Geometrical optics approximation: Geodesics of light rays}
Once an effective optical spacetime is established, other phenomena like bending of light in the black-hole spacetime, must directly follow.
Let us start with the eikonal substitution $\psi=e^{S(r,t)}$ to the wavefunction in eq (15).After some simplification
 we get 
\begin{equation}
-\sqrt{\bar{\varepsilon}^{\theta\theta}}(\frac{\partial S}{\partial t})^2+\frac{1}{\sqrt{\bar{\varepsilon}^{\theta\theta}}}(\frac{\partial S}{\partial r})^2- P(r)=0
\end{equation}
Except for the effective potential term, this is similar to the Hamilton-Jacobi equation for light in curved spacetime
\begin{equation}
g^{00}(\frac{\partial S}{\partial t})^2+g^{rr}(\frac{\partial S}{\partial r})^2+ \frac{1}{r^2}(\frac{\partial S}{\partial \phi})^2=0
\end{equation}

Taking the specific case of the Schwarzschild metric, with the same approximations as discussed before we get
\begin{equation}
-(1-\frac{L}{r})^{-1}(\frac{\partial S}{\partial t})^2+(1-\frac{L}{r})(\frac{\partial S}{\partial r})^2+\frac{J^2}{r^2}=0
\end{equation}
Here ‘J’ is the angular momentum. The similarity in the Hamilton-Jacobi equations immediately implies similar geodesics. The equation of the trajectory in such a spherically symmetric spacetime is given by 
\begin{equation}
\phi=\phi_0 + \int_{\frac{J}{r_1}}^{\frac{J}{r}}{\frac{Jdr}{r^2 \sqrt{\frac{C_0}{g_{00}g_{rr}}-\frac{J^2}{r^2 g_{rr}}}}}
\end{equation}
In the effective optical medium the metric components are replaced by respective functions of permittivity, which act as effective metric.

\subsection{\label{sec:level2}The optical ‘Black Hole’ of Narimanov and Khildishev.}
In order to simulate black-hole phenomena, we must have (a) a medium which simulates the  black-hole spacetime and (b) a perfect absorber, which acts like a horizon, absorbing the entire incident light. We have already developed the formalism for designing the equivalent optical spacetime around a black hole. 
The omnidirectional, broadband optical absorber of Narimanov and Kildishev\cite{narim} can be used to mimic the black hole.
Their model consists of:
1. A central core of complex permittivity, which absorbs all the incident light.
2. An outer shell , with some permittivity profile, which bends the incident light, guiding it to the core and thus ensuring that the entire incident light converges to the core, without any loss.
The model with the core and the shell is placed in an external medium.
Our formalism can be used to arrive at the Hamiltonian used by Narimanov and Kildishev to obtain a Keplerian power law variation of permittivity which ensures proper convergence.  Starting from our eikonal equation (18)  and assuming $S=-Et+$function of $r,\theta,\phi$ and ignoring the derivatives of permittivity we get
\begin{equation}
\frac{1}{\bar{\varepsilon}^{\theta\theta}}(\frac{\partial S}{\partial r})^2+\frac{J^2}{r^2 \bar{\varepsilon}^{\theta\theta}}=E^2
\end{equation}
In \cite{narim} this is expressed as
\begin{equation}
H=\frac{1}{\bar{\varepsilon}^{\theta\theta}}(p_r)^2+\frac{J^2}{r^2 \bar{\varepsilon}^{\theta\theta}}
\end{equation}
(Note that the authors in \cite{narim} use "m" instead of "J" for the angular momentum.)
This gives a ray trajectory  
\begin{equation}
\phi=\phi_0 + \int_{\frac{J}{r_1}}^{\frac{J}{r}}{\frac{d\xi}{\sqrt{C_0 \bar{\varepsilon}^{\theta\theta}(\frac{J}{\xi})- \xi^2 } }} ,\xi=\frac{J}{r}
\end{equation}
The permittivity choice $\bar{\varepsilon}^{\theta\theta}= \bar{\varepsilon_0}^{\theta\theta}[1+(\frac{R}{r})^n]$ for $n\ge2$, in the outer shell will enable maximum convergence and correspondingly facilitates maximum absorption.

This concludes our discussion of the analog black-hole formalism and its application to spherically symmetric spacetimes. Our further investigations will include extending the formalism to the developing of analogues of axially symmetric spacetimes and studying physical effects in their background.
\begin{acknowledgements}
One of us (SSH) would like to thank BASE (Bangalore Association for Science Education) for providing a Research Assistantship and thereby facilitating the completion of a major part of this work.
\end{acknowledgements}

\bibliographystyle{apsrev}
\bibliography{PRDpaper}% Produces the bibliography via BibTeX.

\end{document}